\documentclass{llncs}

\usepackage{xspace}
\usepackage{graphicx}
\usepackage[hyphens]{url}
\usepackage{enumitem}
\usepackage{booktabs}
\usepackage{multirow}
\usepackage[binary-units=true]{siunitx}
\usepackage{subcaption}
\usepackage{color}
\dblfloatsep=\floatsep
\dbltextfloatsep=\textfloatsep
\DeclareMathAlphabet{\mathsc}{OT1}{cmr}{m}{sc}

\newcommand\sarah[1]{}

\newcommand*\dash{\ifvmode\quitvmode\else\unskip\kern.16667em\fi---%
\hskip.16667em\relax}

\newcommand{\hash}{\mathsf{hash}}
\newcommand{\dir}{\mathsf{dir}}

\begin{document}

\title{Why is a Ravencoin Like a TokenDesk?\\ An Exploration of Code Diversity in the Cryptocurrency Landscape}

\author{Pierre Reibel\inst{1}, Haaroon Yousaf\inst{1}, and Sarah
Meiklejohn\inst{1}}

\institute{University College London\\
\email{\{pierre.reibel.16,h.yousaf,s.meiklejohn\}@ucl.ac.uk}
}

\maketitle

\begin{abstract}

Interest in cryptocurrencies has skyrocketed since their introduction a decade
ago, with hundreds of billions of dollars now invested across a landscape of
thousands of different cryptocurrencies.  While there is significant
diversity, there is also a significant number of scams as people seek to
exploit the current popularity.  In this paper, we seek to identify the extent
of innovation in the cryptocurrency landscape using the open-source
repositories associated with each one.  Among other findings, we observe that
while many cryptocurrencies are largely unchanged copies of Bitcoin, the use
of Ethereum as a platform has enabled the deployment of cryptocurrencies with
more diverse functionalities.

\end{abstract}

\section{Introduction}

Since the introduction of Bitcoin in 2008~\cite{satoshi-bitcoin} and its
deployment in January 2009, cryptocurrencies have become increasingly popular
and subject to increasing amounts of hype and speculation.  Initially, the
promise behind cryptocurrencies like Bitcoin was the ability to send
frictionless global payments: anyone in the world could act as a peer in
Bitcoin's peer-to-peer network and broadcast a transaction that\dash without
having to pay exorbitant fees\dash would send money to anyone else in
the world, regardless of their location, citizenship, or what bank they used.
This is achieved by the decentralization inherent in the open consensus
protocol, known as proof-of-work, that allows any peer to not only broadcast
transactions but also act to seal them into the official ledger.

While the realities of Bitcoin have shifted in the ensuing years, the
landscape of cryptocurrencies has also shifted considerably.  There are now
thousands of alternative cryptocurrencies, supporting more exotic
functionalities than the simple atomic transfer of money supported by Bitcoin.
Ethereum, for example, promises to act as a distributed consensus computer
(the Ethereum Virtual Machine, or EVM for short) by enabling arbitrary
stateful programs to be executed by transactions, while Monero and Zcash
promise to improve on the anonymity achieved by Bitcoin transactions.  Others
don't promise new functionalities but instead aim to support the same
functionality as Bitcoin in more cost-effective ways; e.g.,
Zilliqa~\cite{bitcoin-cosi,bitcoin-ng,SP:STVWJG16,CCS:LNZBGS16,omniledger}
and Cardano~\cite{C:KRDO17,EC:DGKR18} incorporate respective ideas from the
academic literature about achieving consensus without relying entirely on
proof-of-work.

Alongside this rapid expansion in the functionality of cryptocurrencies
(or indeed the general applicability of the underlying concept of a
blockchain), there has also been a genuine explosion of investment into these
technologies.  In July 2013, for example, there were 42 cryptocurrencies
listed on the popular data tracker
CoinMarketCap,\footnote{\url{https://coinmarketcap.com/historical/20130721/}}
and the collective market capitalization was just over 1 billion USD.  In
July 2018, in contrast, there were 1664 cryptocurrencies, and the collective
market capitalization was close to 1 trillion USD.  While comprehensive in terms
of deployed cryptocurrencies, this list does not even include many of the recent
``initial coin offerings'' (ICOs) that have similarly attracted
millions in investment despite there having been many documented
scams.\footnote{\url{https://deadcoins.com/}}\footnote{\url{https://magoo.github.io/Blockchain-Graveyard/}}
Against this backdrop of hype and investment, it is thus crucial to gain some
insight into the different types of functionalities offered by these many
different cryptocurrencies, to understand which coins offer truly novel
features and are backed by genuine development efforts, and which ones are
merely hoping to cash in on the hype.

This paper takes a first step in this direction, by examining the entire
landscape of cryptocurrencies in terms of the publicly available source code
used to support each one.  While source code may not be the most
accurate representation of a cryptocurrency (as, for example, the actual
client may use a different codebase), it does reflect the best practices of
the open-source software community, so we believe it to be a reasonable proxy
for how a cryptocurrency does (or should) represent itself.
In particular, we begin by describing our collected
dataset in Section~\ref{sec:data}, where we can already observe that many listed
cryptocurrencies are in fact tokens based on the Ethereum blockchain.  We then
move on in Section~\ref{sec:forks} to a general identification of the ways in
which cryptocurrencies copy their ideas from those of others, and
observe that Bitcoin is by far the most popular in this respect.  Due to the
dominance of Bitcoin and Ethereum, we explore them in more detail in
Sections~\ref{sec:bitcoin}
and~\ref{sec:ethereum} before concluding in Section~\ref{sec:conclusions}.

\section{Related Work}\label{sec:related}

We treat as related research that measures either general properties of
open-source software, or research that measures properties of
cryptocurrencies.
In terms of the former, there have been numerous papers measuring
GitHub repositories.  For example, Hu et
al.~\cite{hu2016influence} and Thung et al.~\cite{thung2013github} measured
the influence of software projects according to their position of their
repositories and developers in the GitHub social graph, and others have
taken advantage of the volume of source code available on GitHub to analyze
common coding practices~\cite{github-folders} or how bugs vary across
different programming languages~\cite{ray2014github}.

In terms of the latter, there are by now many papers that have focused on
measuring properties of both the peer-to-peer
networks~\cite{FC:KosKosMcD14,FCW:DonPerHer14,CCS:BirKhoPus14,SP:ApoZohVan17}
and the blockchain data associated with
cryptocurrencies~\cite{reid2013analysis,FC:RonSha13,meiklejohn-fistful,FC:SpaMagZan14,miller2017empirical,sarah-zcash,FC:BHHMNW14,FC:VBCKM16}, as well as their
broader ecosystem of participants~\cite{FC:MooChr13,FCW:VasThoMoo14,FC:VasMoo15,marie-ponzi}.
Given the volume of research, we focus only on those papers most related to
our own, in that
they analyze properties across multiple cryptocurrencies, rather than within a
single one like Bitcoin.
In terms of comparing Bitcoin and Ethereum, Gencer et
al.~\cite{FC:GBERS18}
compared the level of decentralization in their peer-to-peer networks and
found, for example, that Ethereum mining was more centralized than it was in
Bitcoin, but that Bitcoin nodes formed more geographic clusters.
Azouvi et al.~\cite{FCW:AzoMalMei18} also compared their level of
decentralization, in terms of the discussions on and contributions to
their GitHub repositories, and found that Ethereum was more centralized in
terms of code contribution and both were fairly centralized in terms of the
discussions.
Gervais et
al.~\cite{CCS:GKWGRC16} introduced a framework for identifying the tradeoff
between security and performance in any cryptocurrency based on proof-of-work,
and found that the same level of resilience to double-spending attacks was
achieved by 37 blocks in Ethereum as by 6 blocks in Bitcoin.  Finally, Huang
et al.~\cite{FC:HuaLevSno18} compared the effectiveness of
different mining and speculation activities for 18 cryptocurrencies, and found
that the profitability of both was affected by when a cryptocurrency was
listed on an exchange.

\section{Background}\label{sec:back}

Here we provide some brief background on the functionality of Bitcoin and
Ethereum, which are the two biggest cryptocurrencies in terms of their market
capitalization.
As general terminology, we use \emph{cryptocurrency} to refer to anything
with an exchangeable
unit of value, \emph{coin} to refer to a cryptocurrency with its own dedicated
blockchain, and \emph{token} to refer to a cryptocurrency that operates instead
using the blockchain of another cryptocurrency (e.g., Ethereum).

Fundamentally, the main functionality offered by Bitcoin is the atomic
transfer of money from a sender (or set of senders) to a recipient (or set of
recipients).  This is supported by a limited scripting language, known simply
by the name Script, which defines how transactions are created and verified.
Bitcoin is also, however, a standalone platform, and thus its codebase must do
significantly more than support this so-called transaction
layer~\cite{SP:BMCNKF15}.  In particular, it must define the peer-to-peer
network, by which clients can find and communicate with each other, and the
consensus protocol, by which they can come to agreement on the transactions
that have taken place.

Beyond the relatively simple functionality offered by Bitcoin, Ethereum allows
developers to create and deploy \emph{smart contracts} onto the blockchain.
These are stateful programs, typically written in a language called Solidity,
that can be triggered by transactions and used to
run (almost) arbitrary code.  The only limitation is their complexity, as
every operation they perform has an associated cost, and transactions
have a maximum amount they are allowed to spend.

There are two special types of smart contracts, ERC20 and ERC721, that define
\emph{tokens}: ERC20
tokens\footnote{\url{https://github.com/ethereum/eips/issues/20}}
are designed to be currency, and thus are fungible, whereas ERC721 tokens may
be non-fungible and thus support collectibles such as
CryptoKitties.\footnote{\url{https://www.cryptokitties.co/}}  At the
most basic level, a token contract is a ledger mapping owners
(identified by their Ethereum address) to the amount of tokens that they own,
along with an associated set of rules determining how tokens are transferred
between owners.

\section{Data Collection}\label{sec:data}

In order to collect the source code associated with each cryptocurrency, we
started with the list maintained at
CoinMarketCap,%
which is generally
regarded as one of the most comprehensive resources for cryptocurrency market
data.  The site maintains not only market data for each cryptocurrency (its
market capitalization, price, circulating supply, etc.), however,
but also links to any websites, blockchain explorers, or\dash crucially for
us\dash source code repositories.  We last scraped the site on July 24 2018,
at which point there were 1664 cryptocurrencies listed, with a cumulative
market capitalization of 293B USD.

\begin{table}[t]
\centering
\begin{tabular}{l@{\hskip 8pt}c@{\hskip 8pt}l}
\toprule
Category & \# coins & Representative examples \\
\midrule
Animals & 29 & RabbitCoin, Birds \\
Computing & 47 & AI Doctor, Decentralized Machine Learning \\
Drugs & 13 & Cannation, KushCoin \\
Finance & 22 & iBank, Intelligent Investment Chain \\
Food & 10 & EggCoin, Olive \\
Gambling & 17 & CashBet Coin, CasinoCoin \\
Geography & 12 & Asiadigicoin, NewYorkCoin \\
Nation states & 37 & PutinCoin, Shekel, BritCoin \\
Outer space & 17 & Marscoin, SpaceChain \\
Metals \& precious stones & 37 & GoldPieces, PlatinumBAR, Rubycoin \\
Religion & 7 & BiblePay, HalalChain \\
\bottomrule
\end{tabular}
\caption{Different categories of cryptocurrencies, based on the name of the
coin, along with some representative examples.}
\label{tab:categories}
\hrulefill
\end{table}

Of these cryptocurrencies, there were 866 categorized as a token
rather than as a coin.
There were 366 cryptocurrencies with a stated price of \$0.00, and in fact
1468 (88\%) had a stated price of less than \$1.00.  There were
276 cryptocurrencies with an unknown circulating supply (and thus an unknown
market capitalization), and 924 (55.5\%) had a market capitalization of over
1M USD.  The names of the cryptocurrencies were typically designed to evoke a
specific concept; e.g., wealth, computing, politicians, or existing
fiat currencies.
Based on their names, we manually sorted
all of the listed cryptocurrencies into the categories shown in
Table~\ref{tab:categories}.

\subsection{Source code repositories}\label{sec:repos}

Of the listed cryptocurrencies, 1123 had a link available to some source code
repository.  We performed manual spot checking to ensure that the links were
legitimate, and in some
cases replaced the links where the information was inaccurate (for Bitcoin
Cash, for example, the provided link was for the repositories backing
\url{bitcoincash.org} rather than the actual software code).  Of these, 1108
pointed to GitHub (98.7\%).

As should be expected, many of the cryptocurrencies had multiple software
repositories available; indeed, the links provided were to the lists of
repositories for a given GitHub organization, and in total there were 13,694
individual repositories available.  The vast majority of these repositories
had been created after October 2014, with a notable rise in frequency
starting in April 2017.  %
These repositories typically fell into one of three categories: (1) integral
to the cryptocurrency itself, such as
implementations of the reference client or supporting libraries; (2)
irrelevant, such as a different project by the same organization; or (3)
unchanged forks or mirrors of popular software projects, such as llvm.
Given our goal of differentiating between different
cryptocurrencies, we sought to isolate the first category of
``meaningful'' code.

Our initial hypothesis was that more integral repositories
would be (1) better maintained, (2) more popular, and (3) re-used more
frequently.
There are several open-source tools for determining the ``health'' of the
maintenance of a GitHub repository~\cite{repo-health}, which typically
measure the activity, level of contribution, and responsiveness to pull
requests and issues.  In fact, there is even a tool called CoinCheckup that
does this specifically for
cryptocurrencies.\footnote{\url{https://coincheckup.com/analysis/github}}
We observed, however, that many of the repositories had a fairly low level
of activity, so this was not on its own a good way to distinguish between
different repositories.
Another natural measure would be the total number of commits to a repository,
but this information is not readily accessible via the GitHub API.\footnote{It
    is possible to derive it from information given for individual
    repositories, albeit in a relatively inefficient manner, but not from an
    organization's list of repositories.}
We thus measured activity according to the gap in time (in weeks)
between the current time and the last update of the repository, under the
assumption that this would be shorter for more active repositories.

In terms of the second two criteria, the most natural
representation of popularity and reusability is the number of stars and
forks~\cite{github-stars-forks}.  We again found, however, that stars
were not a useful indication of whether or not a repository
contained relevant source code.  For example, important meta-information
related to a cryptocurrency such as its whitepaper or its improvement
proposals (IPs) was often contained in highly-starred repositories.  We thus
chose to ignore stars and look solely at forks.

Finally, in terms of relevance, we
favored repositories with a name identical or close to that of the
cryptocurrency, as well as ones that indicated they contained source
code (e.g., with `core' in the name) or code otherwise relevant to the
operation of the cryptocurrency (e.g., `token' or `contract' for ERC20 tokens).
We also excluded repositories whose names contained terms that indicated they
were not relevant; e.g., `website' or `docs.'  A complete
list of these excluded terms can be found in Table~\ref{tab:bad-terms} in
Appendix~\ref{sec:extras}.

\newcommand\exclusiontable{
\begin{table}[ht]
\centering
\begin{tabular}{lc}
\toprule
Concepts & Excluded terms \\
\midrule
Binaries & binaries \\
Documentation & docs, document, papers, whitepaper, wiki \\
Improvement proposals (IPs) & \texttt{\^[a-z]{1,2}ips} \\
Other & electrum, explorer, faucet, vanitygen \\
Packaging & docker, homebrew, install \\
Testing & example, test \\
Tools & kit, lib, plugin, sdk, service, tools \\
User interface & android, gui, ios, macos, mobile, window \\
Website & .com, .info, .io, .net, .org, -org, site, website, www \\
\bottomrule
\end{tabular}
\caption{The terms which, if we encountered them in the name of a repository,
led to the exclusion of that repository from further consideration.  We added
manual exceptions to these where appropriate (e.g., allowing the `ips' pattern
for the CHIPS and Vipstar Coin cryptocurrencies).}
\label{tab:bad-terms}
\hrulefill
\end{table}
}

In the end, we assigned a rating to each repository for a given
cryptocurrency according to: (1) the gap between its last update and the
current date, to capture activity (where this was subtracted from the rating,
as a longer gap indicates less activity); (2) its number of forks, to capture
popularity and reuse; and (3) information about the name of the repository,
to capture relevance.  For each cryptocurrency, we then cloned the top 20\% of
the list of repositories, sorted from high to low by these ratings (or
cloned one repository, whichever was larger).  Even after
compiling this list, we made various manual adjustments in order to ensure
that we had selected the ``right'' repositories.  %
We cloned 2354 repositories in total, which
comprised roughly 100~GB of data.

\subsection{Deployed source code}\label{sec:contracts}

As evidenced by the 866 (52\%) listed cryptocurrencies that were categorized
as tokens (and the fact that 74 of these even had `token' in their name), it
is popular to launch new cryptocurrencies not as standalone
coins, but as tokens that are supported by
existing cryptocurrencies.  Of these, by far the most popular type is an ERC20
token, supported by Ethereum.  %
Of these listed tokens, 406 did not have any source code link available.  For
ERC20 tokens that have been deployed, however, it is often possible to obtain
the contract code from another source: the version deployed on the
Ethereum blockchain itself is compiled bytecode, but it is common practice to
provide the Solidity code and display it on blockchain explorers
such as Etherscan.\footnote{\url{https://etherscan.io/}}

For these tokens, we thus chose to use Etherscan as a data source (in
addition to any provided repositories), in order to aid our Ethereum-based
analysis in Section~\ref{sec:ethereum}.  At the time that we scraped
Etherscan, there were 612 ERC20 tokens listed, identified by a name and a
currency symbol (e.g., OmiseGO and OMG).  Of these, we found 438 with a match
on CoinMarketCap, where we defined a match as having (1) identical currency
symbols, and (2) closely matching
names. (We couldn't also require the name to be identical
because in some cases the name of the contract was somewhat
altered from the name of the cryptocurrency; e.g.,
SPANK instead of SpankChain.)  %
We scraped the available contract code for each of these tokens, which in all
but 9 cases was Solidity code rather than just on-chain bytecode.  We thus
ended up with 429 deployed ERC20 contracts.

\section{Detecting Code Reuse}\label{sec:forks}

In this section,
we attempt to identify the extent to which cryptocurrencies reuse the codebases
of others, in order to identify which cryptocurrencies incorporate
genuinely novel ideas and which ones largely derive their ideas from
others.  We do so using several different
techniques, ranging from very simplistic (seeing if the name of one is a
prefix of the name of another) to more complex.  We refer to
the cryptocurrencies borrowing ideas from others as \emph{derivatives}.

Most of our methods label one cryptocurrency as a derivative of another if it
has at least one repository that appears as derived from a repository of the
second cryptocurrency.  This means that,
for example, if a cryptocurrency copied the Bitcoin repository but its actual
platform consists largely of repositories
written from scratch, we may unfairly label it as a derivative of Bitcoin.  On
the other hand, if we tried to do otherwise then we might unfairly reward
cryptocurrencies that copy the Bitcoin repository and then create many other
repositories that do not contribute to the functioning of the
platform (e.g., web templates).  Ultimately, without significantly more
advanced analysis to determine how repositories are linked and which
ones meaningfully support the platform, this is a
potential limitation that we must accept in our analysis that follows.

\subsection{Name derivations}\label{sec:name-forks}

As a first simple method, we observed that many
cryptocurrencies attempt to profit from the name recognition of popular
cryptocurrencies by using a similar name; e.g., Bitcoin Planet or Ethereum Gold.
We thus decided to identify one cryptocurrency as a derivative of another
cryptocurrency if the name of the second is a prefix of the name of the first
(as in the examples above), manually excluding cryptocurrencies whose names are
common words and thus might be prefixes anyway (e.g., Crypto).
Using this
method, we identified 28 cryptocurrencies with one derivative, and 8 with two
derivatives (0x, Aurora, Dynamic, Hyper, Litecoin, Monero, Sentinel, and
Waves).
While they did not exactly match our prefix method,
there were four cryptocurrencies whose names
seem designed to evoke the ideas behind the popular ``privacy coin''
Zcash: ZClassic,
ZCoin, Zoin, and
Zero.  Finally, the two most derived cryptocurrencies are\dash
unsurprisingly\dash Bitcoin (with 17 derivatives) and Ethereum (with 6).
While this method already yields some interesting insights into the extent to
which the Bitcoin brand has been borrowed, the name
of a cryptocurrency is not necessarily indicative of the contents of its
underlying codebase.  We thus continue to develop more
code-specific methods for determining derivatives.

\subsection{Git forks}\label{sec:commit-forks}

\begin{figure}[t]
\centering
\includegraphics[width=0.74\textwidth]{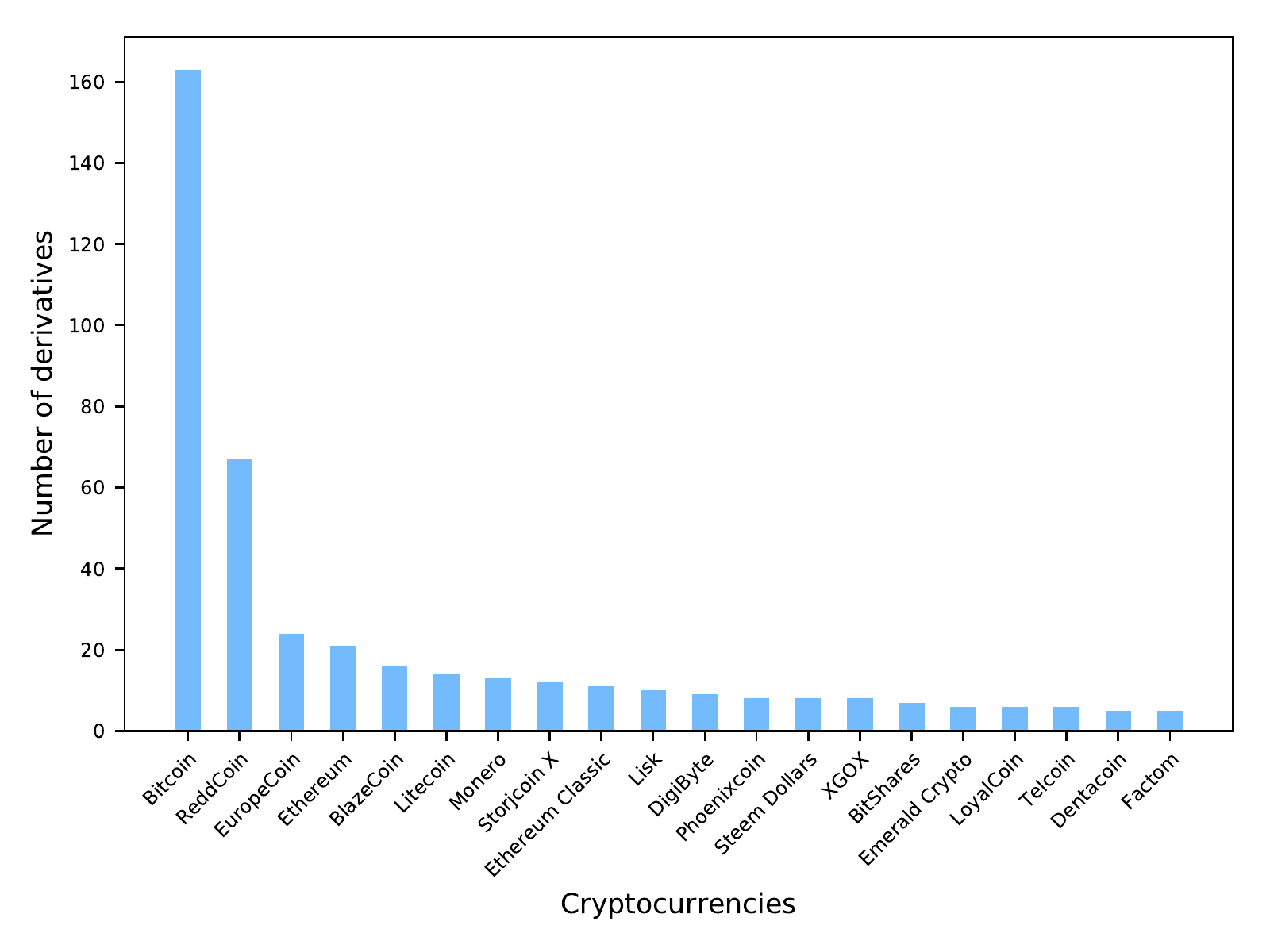}
\caption{The cryptocurrencies with three or more derivatives, where a
derivative is defined as a cryptocurrency that has a repository initially
created by the first one.}
\label{fig:git-forks}
\hrulefill
\end{figure}

Next, we considered the Git forks of a cryptocurrency, meaning the
cryptocurrencies that were created as a fork of the GitHub repository of
another cryptocurrency.
To do this, we identified (by hash) every commit to every repository we
cloned.  We then mapped commits to the lists of the repositories in
which they appeared, and considered the oldest repository containing that
commit to be the original cryptocurrency and all
other repositories in the list to be derivatives.  To elevate this
to the level of cryptocurrencies,%
we labelled a cryptocurrency as a derivative of another one
if any of its repositories were forked from any repositories of the
second one.  The results are in Figure~\ref{fig:git-forks}.
The most immediate observation is that this method captures
significantly more derivatives than the name-based one, and that it reinforces
the popularity of Bitcoin
and Ethereum (according to this method, Bitcoin has 163 forks and Ethereum has
21), as well as of Litecoin and Monero.

There are also, however, several cryptocurrencies with many Git forks that are
less well-known; e.g., Reddcoin and EuropeCoin.  These are due to being the
earliest cryptocurrencies to incorporate independently popular repositories,
and thus highlight
the main limitation of this method; namely, that the earliest cryptocurrency
to fork a popular general-purpose library or other integral software
development tool will be incorrectly labelled as popular.  In addition, it
doesn't consider ``intermediate'' derivatives; e.g., Litecoin and Peercoin are
both forks of Bitcoin that have themselves been forked many times.

To get a sense of how much this method underestimated code
reuse, we considered the file \verb#addrman.cpp#, which is the way
addresses are managed in Bitcoin's peer-to-peer network and is one of the most
reused cryptocurrency-specific files we saw.  It appeared in a repository
for 536 other cryptocurrencies, meaning there are many potential derivatives of
Bitcoin that this method does not capture.

\subsection{Copyright derivations}\label{sec:copyright}

Most open-source repositories are not written from scratch, but instead make
use of established libraries or other code.  This is expected (and in fact
encouraged), as long as the authors acknowledge the original
authors.  We thus
decided next to use this copyright information in order to identify potential
derivatives.

More specifically, we looked in every repository to see if any COPYING files
were available.  If they were, then we scraped the comments concerning
copyright from the beginning of these files.  If not, then we went through every
source code file in the repository and scraped the copyright information there
instead, where we identified source code according to the file extensions
maintained by the CLOC (Count Lines of Code)
library.\footnote{\url{https://github.com/AlDanial/cloc}}
Either way, we then considered the collective authors of the repository to be
the union of all of the individual authors identified by the copyright lines.
\sarah{Some of this may change if we decide to look for bad guys here too.}

Given this set of collective authors, our next task was to assign
them to a specific cryptocurrency.  Luckily, many of the cryptocurrencies do
not identify contributors by their individual names, but rather by the name
of the cryptocurrency; e.g., ``Bitcoin Developers'' or ``The go-ethereum
Authors''.  To handle the several prominent exceptions to this rule, we manually
created a mapping from popular individual contributors to the coins with which
they were associated; e.g., Pieter Wuille for Bitcoin.
This also covered cases in which the name
of the coin was altered slightly (e.g., PPCoin
for Peercoin) and in which the copyright was given to the
organization rather than the cryptocurrency (e.g., IOHK for Cardano).  We did
not include contributors to popular open-source libraries (e.g., Boost
and LevelDB) in order to keep our focus on cryptocurrency-specific code rather
than standard software development tools.

\begin{figure}[t]
\centering
\includegraphics[width=0.74\textwidth]{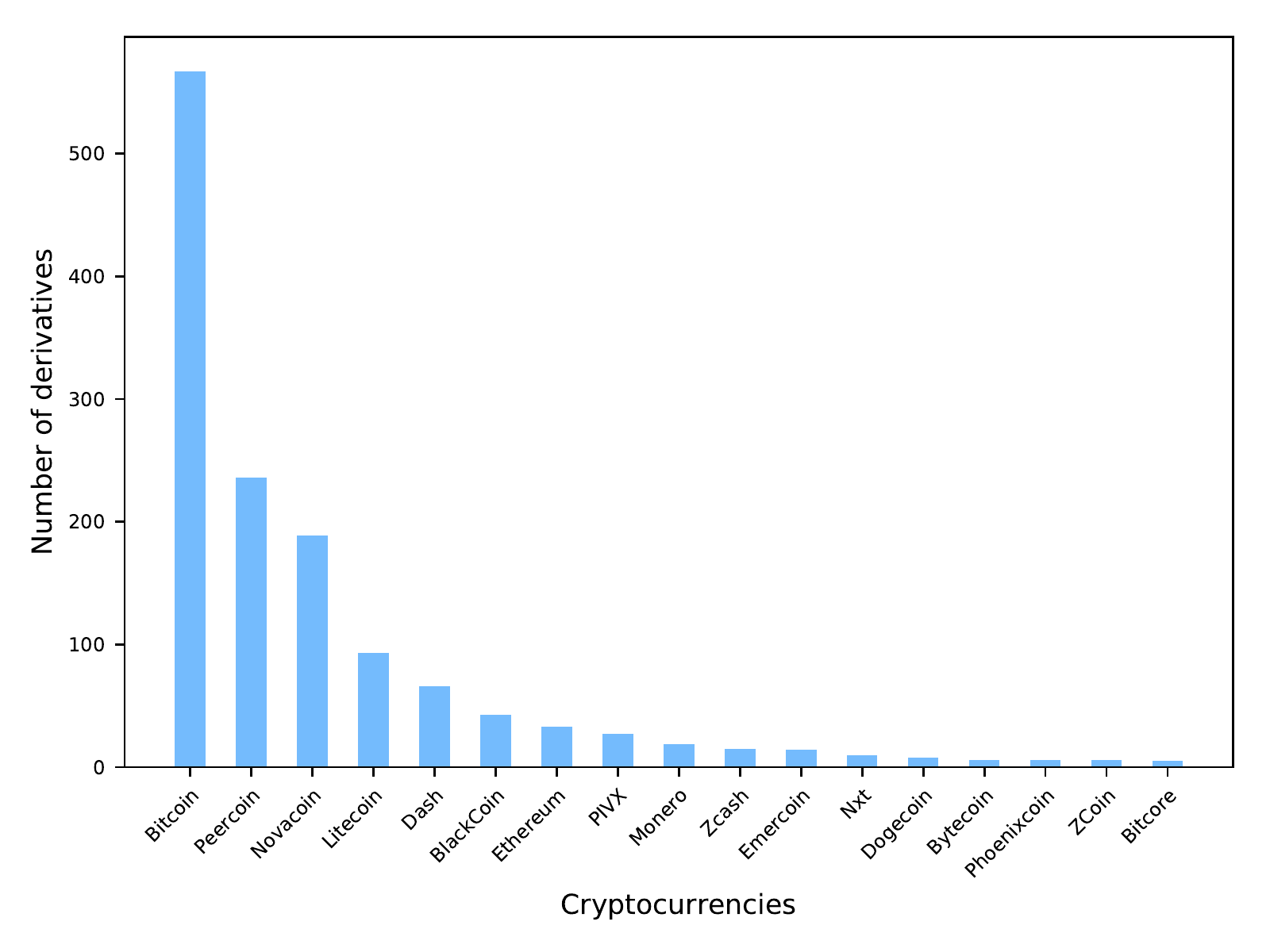}
\caption{The cryptocurrencies with five or more derivatives, where a
derivative is defined as a cryptocurrency that contains information
copyrighted by the first one.}
\label{fig:copyright-forks}
\hrulefill
\end{figure}

With this information, we then labelled one cryptocurrency as a
derivative of another if the copyright lines in the first included the name of
the second (or the name of one of its popular contributors).  The results are
in Figure~\ref{fig:copyright-forks}.
This graph again demonstrates the dominance of Bitcoin, as well as of several
cryptocurrencies (most notably Litecoin, Peercoin, and Novacoin) that have
also been very popular to fork.  It also comes much closer to the expected
number of derivatives given the occurrence of \verb#addrman.cpp# discussed in
Section~\ref{sec:commit-forks}.

\begin{figure}[t]
\centering
\includegraphics[width=0.45\textwidth]{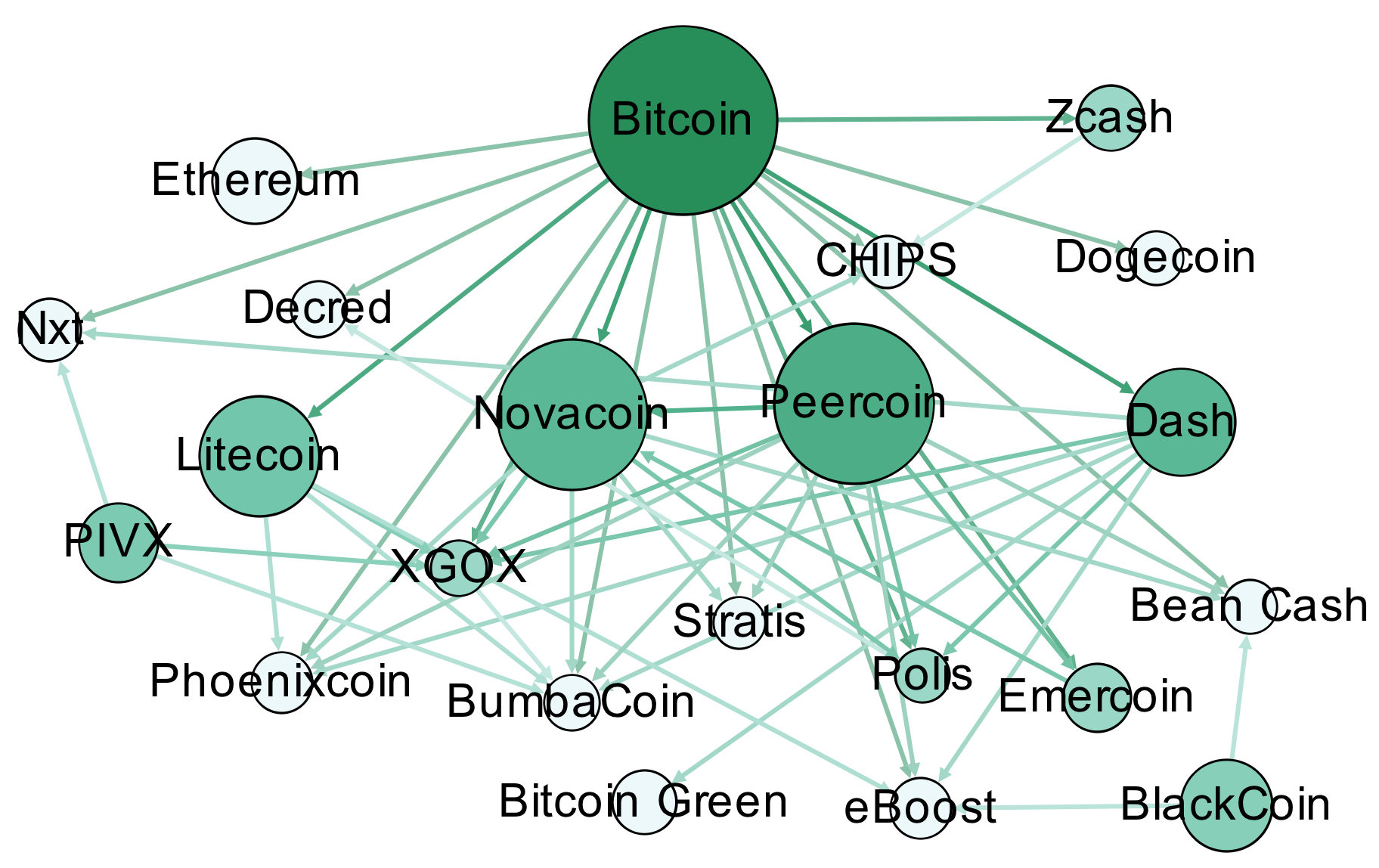}
\caption{In the largest connected component of the graph of copyright
derivations, the cryptocurrencies with degree over 7, where the size (and color)
of the node is proportional to its degree.}
\label{fig:copyright-graph}
\hrulefill
\end{figure}

In addition to capturing a richer type of derivation, this approach also had
the benefit over the previous ones that it did not flatten transitive
relationships; i.e., if X forked from Y, which was itself a fork of Z, then X
would be labelled as a derivative of both Y and Z, rather than just Z.  It
thus builds a tree of derivations, such as that seen on the (currently
unmaintained) website Map of Coins.\footnote{\url{http://mapofcoins.com}}  A
version of this tree with the most derived (or derivative) cryptocurrencies
can be seen in Figure~\ref{fig:copyright-graph}.
This graph reinforces the results from Figure~\ref{fig:copyright-forks}, in
terms of which cryptocurrencies have the most derivatives.  It also highlights
cryptocurrencies like BumbaCoin, which incorporate code from many different
(10) sources.%

While this method nicely captures code reuse, it goes perhaps a bit too far in
labelling derivatives; again, code reuse is encouraged as a form of library
support and using or modifying copyrighted code from another cryptocurrency
does not indicate a lack of other novel ideas.  Thus, our
final method considers not just whether or not a cryptocurrency uses the
codebase of another, but the actual proportion of code that goes unmodified.

\subsection{File derivations}\label{sec:hash}

Finally, we looked for the most direct form of derivation: taking
another repository and using it without any modification.  To identify this,
we computed and stored the hash of every source code file
in our cloned repositories; as in Section~\ref{sec:copyright}, we identified
source code file extensions using the CLOC library.  We then computed a
similarity score $S_\hash$ between a repository $A$ and another one
$B$ by counting the number of files in $A$ with an
identical file in $B$ (meaning the hash was the same), and then dividing by
the total number of files in $A$.  To elevate this to the level of
cryptocurrencies $C_1$ and $C_2$, we then computed $S_\hash(C_1,C_2)$ as
\[
S_\hash(C_1,C_2) = \frac{\sum_{A\in C_1}S_\hash(A,\cup_{B\in C_2} B)}
{\sum_{A\in C_1} \# \textrm{ files in } A};
\]
i.e., for each repository $A$ contributing to $C_1$ we counted
the number of files that were identical to a file in \emph{any} repository
contributing to $C_2$, and then divided this by the
total number of files across all repositories contributing to $C_1$.

\begin{figure}[t]
\centering
\includegraphics[width=0.85\textwidth]{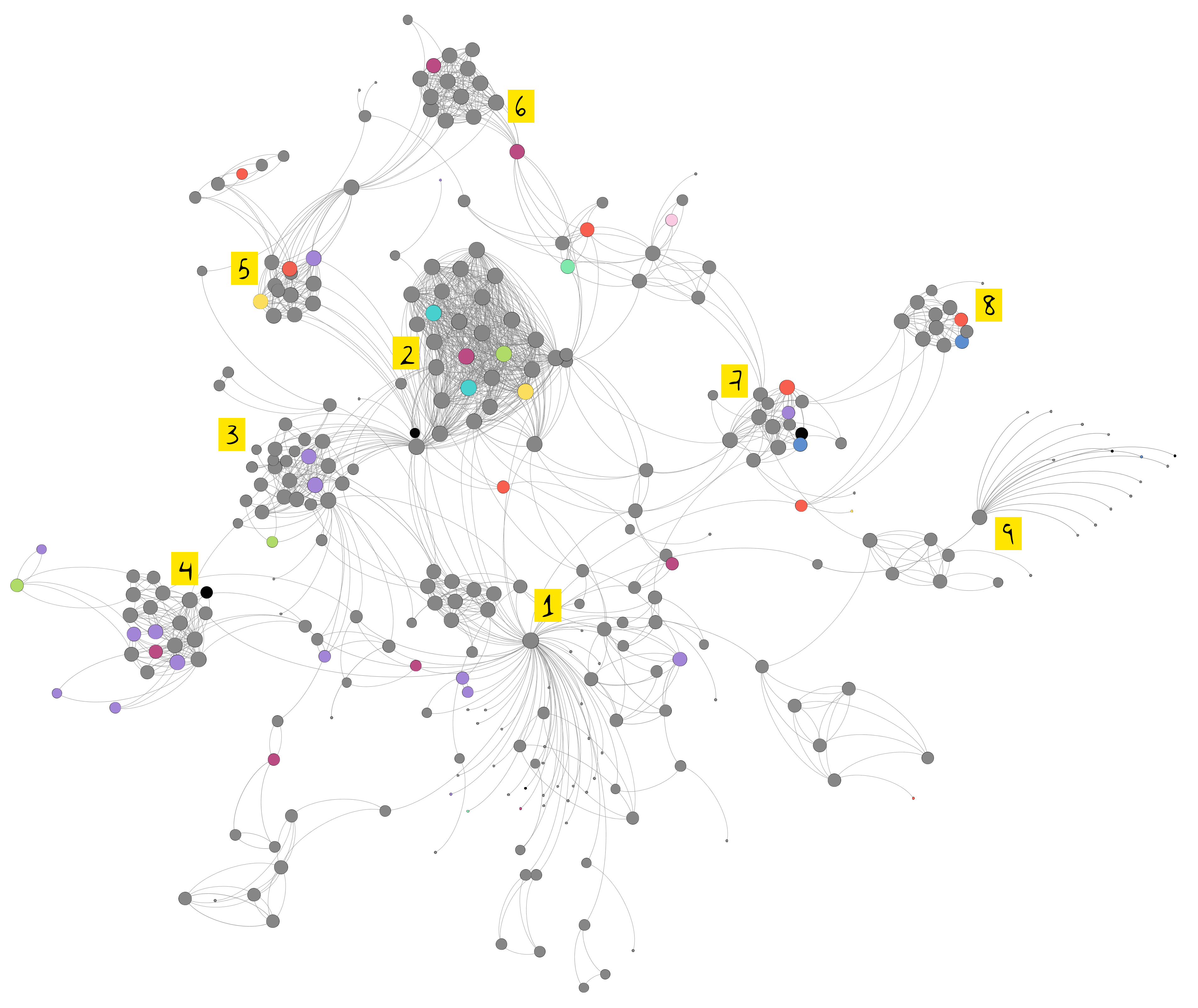}
\caption{The largest connected component of the graph formed by creating an
edge from $A$ to $B$ if $S_\hash(A,B) > 0.7$, along with labels for the most
prominent clusters.  The nodes are colored according to the categories from
Table~\ref{tab:categories}.}
\label{fig:hash-graph}
\hrulefill
\end{figure}

We ran this for every pair of cryptocurrencies $A$ and $B$ (for both
$S_\hash(A,B)$ and $S_\hash(B,A)$, since they are not symmetric), and used the
results to create a graph in which nodes represent cryptocurrencies and there
is a directed edge from $A$ to $B$ if $S_\hash(A,B) > 0.7$.  This resulted in a
graph with 445 nodes and 1854 edges, the largest
connected component of which can be seen in Figure~\ref{fig:hash-graph}
(consisting of 302 nodes and 1599 edges).

Most of this component consists of Bitcoin forks, so we defer further
discussion of these clusters to Section~\ref{sec:bitcoin}.  The exception is
cluster~9, which consists of one cryptocurrency (Zeepin) that is 100\%
similar to 16 other cryptocurrencies.  The reason is simple:
its repository consisted solely of an LGPL-3.0 license, so it
matched other repositories with the same version of this license.  At the time
we scraped CoinMarketCap, Zeepin had a market capitalization of 23 million
USD.

\section{Bitcoin and Its Derivatives}\label{sec:bitcoin}

\begin{figure}[t]
\centering
\includegraphics[width=0.7\textwidth]{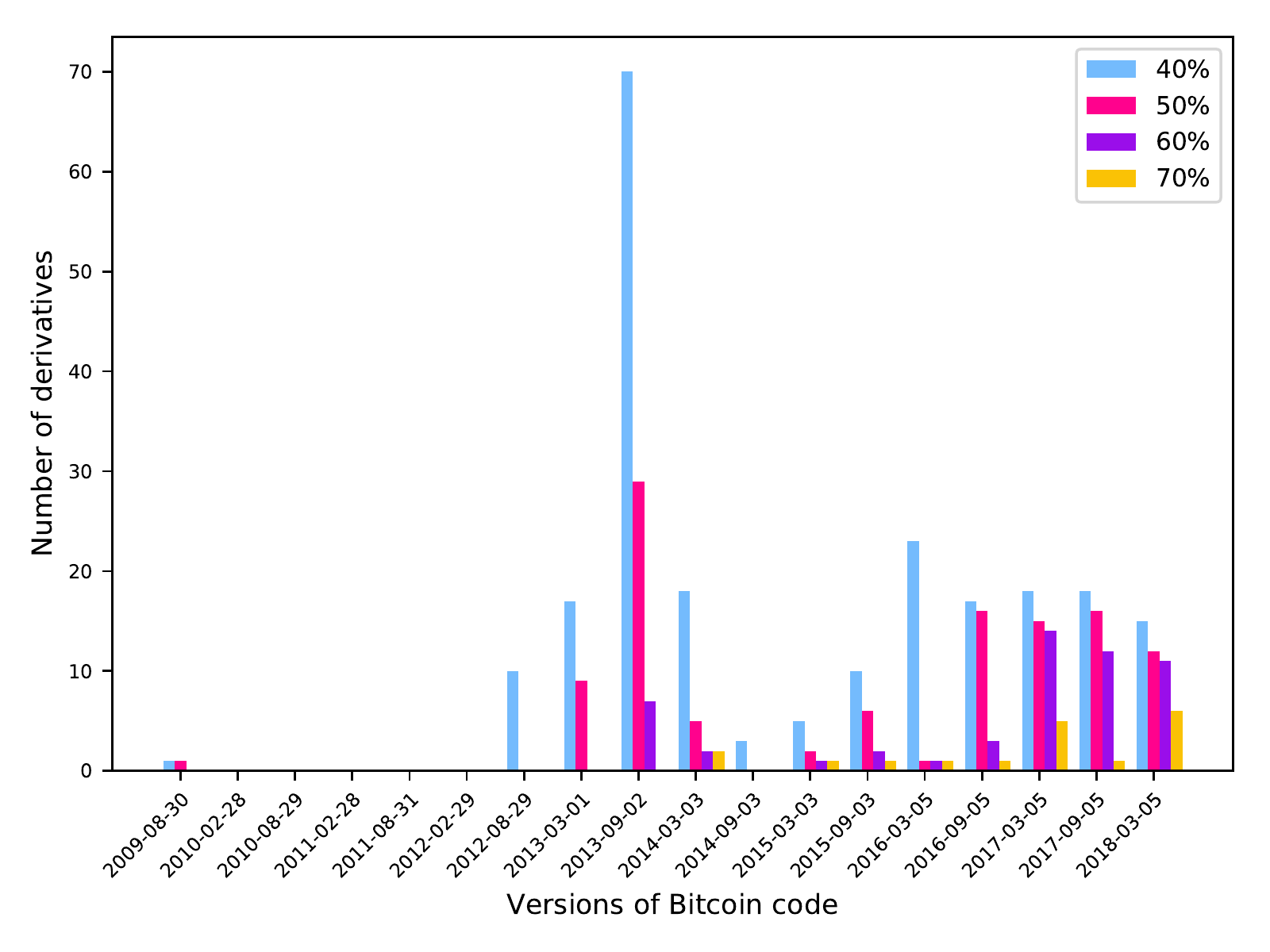}
\caption{For different percentage overlaps, the number of cryptocurrencies
matching a given version of the Bitcoin codebase in terms of the hash
similarity.}  %
\label{fig:bitcoin-hash-sim}
\hrulefill
\end{figure}

In Section~\ref{sec:forks}, we saw a clear dominance of Bitcoin in terms of
the reuse of its code by other cryptocurrencies.
We now explore these Bitcoin forks in more detail, by comparing them not only
against one version of the Bitcoin codebase, but against multiple versions
reflecting its evolution over time.
To do this, we first
created a list of all commits to the Bitcoin repository, starting from August
2009, and collected one version of the codebase every six months after that.
We labelled these versions from 1 to 18, but
stress that these labels are not correlated with any official
releases.  We then re-ran the hash similarity code from
Section~\ref{sec:hash} against each of these versions, assigning a
cryptocurrency to a single version by picking the one with which it had
the highest overlap.
The results are in Figure~\ref{fig:bitcoin-hash-sim}, and demonstrate an
increase in the number of derivatives over time, with a clear spike at
version~9 (which represents the codebase in September 2013).

Hashing is a very brittle method of comparison, however, and what we saw
copied far more often than the exact file contents was Bitcoin's directory
structure.
We thus also compared the directory structure of two repositories; i.e., we
computed a similarity score $S_\dir$ between a repository $A$ and another
one $B$ by counting the number of files in $A$ with an
identical filename in $B$ (meaning the name and path was the same), and then
dividing by the total number of files in $A$.  We elevated this to the level
of cryptocurrencies in the same way as we did in Section~\ref{sec:hash} for
$S_\hash$.  For completeness, the results are in
Figure~\ref{fig:bitcoin-dir-sim} in Appendix~\ref{sec:extras}, and demonstrate
that the directory structure (and in particular the one associated with older
versions of the codebase, prior to its re-organization due to SegWit adoption)
is heavily borrowed by other cryptocurrencies.

\newcommand\bitcoindir{
\begin{figure}[t]
\centering
\includegraphics[width=0.7\textwidth]{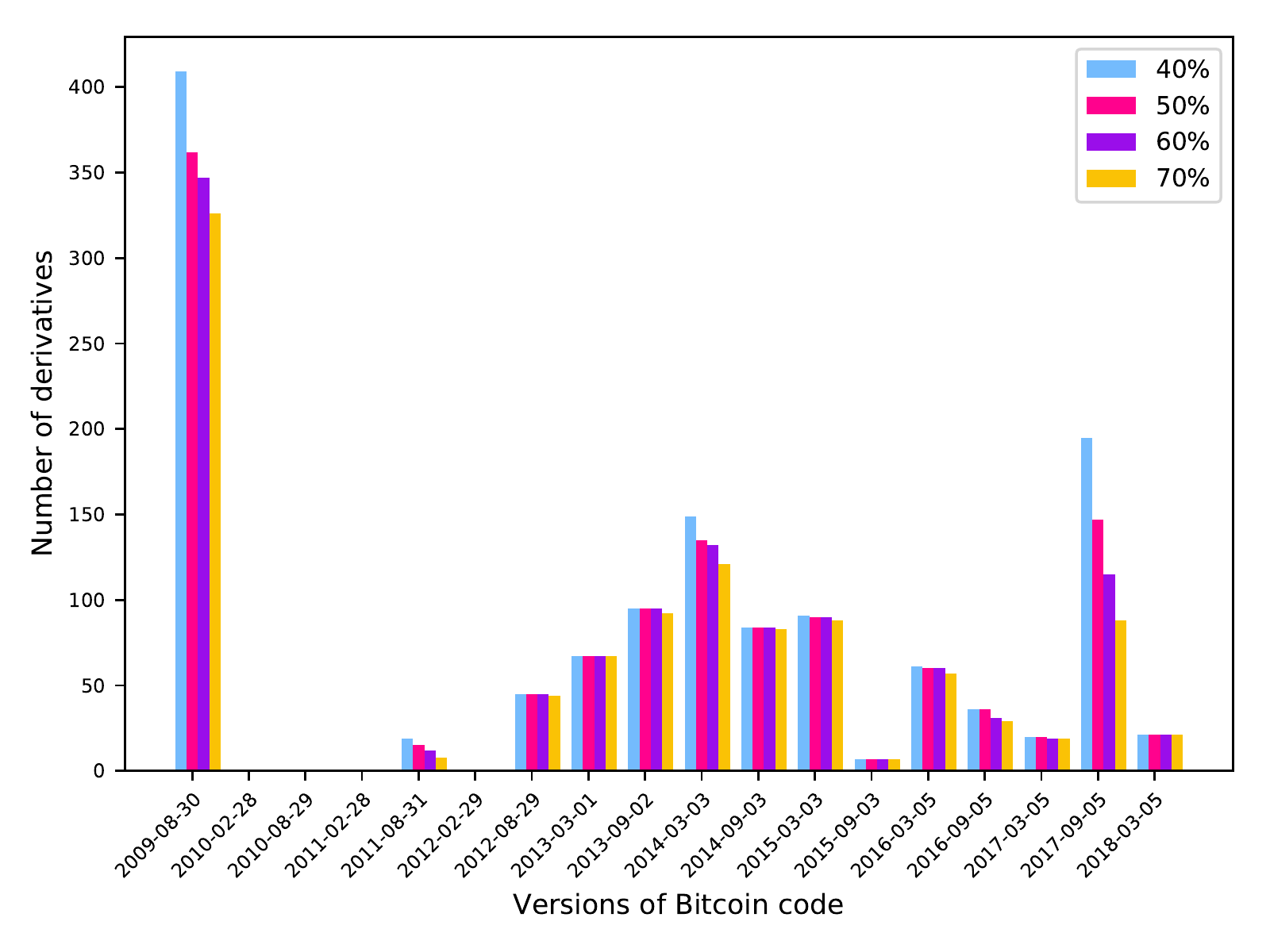}
\caption{For different percentage overlaps, the number of cryptocurrencies
matching a given version of the Bitcoin codebase in terms of the directory
similarity.}
\label{fig:bitcoin-dir-sim}
\hrulefill
\end{figure}
}

Finally, we revisit the graph in Figure~\ref{fig:hash-graph}, as this
connected component is almost entirely associated with Bitcoin forks.  In
particular, we can briefly explain clusters~1-8 as follows:

\begin{itemize}

\item{\textbf{1.}} The node at the center of this
cluster, Akuya Coin, has a directory structure similar (63\%) to a version
of the Bitcoin codebase from 2013, but many (32\%) of its files are empty
and thus have the same hash, which makes it appear similar to 76 other
Bitcoin forks.

\item{\textbf{2 and 3.}} Both of these clusters also have a directory
structure similar to older versions of the Bitcoin codebase (the average
directory similarity was 89\% for cluster~2 and 82\% for cluster~3), and are
similar to the same cryptocurrency (BumbaCoin).  Many
also incorporate the Zerocoin
code:\footnote{\url{https://github.com/Zerocoin/libzerocoin}} 84\% of the
nodes in cluster~2 and 65\% of the nodes in cluster~3.  This is notable given
that this code comes with the emphatic warning ``THIS CODE IS UNMAINTAINED AND
HAS KNOWN EXPLOITS. DO NOT USE IT.''  In total it is included in
repositories for 97 different cryptocurrencies.

\item{\textbf{4 and 5.}} These clusters were the ones most similar to
Bitcoin: on average we had $S_\hash = 0.51$ and $S_\dir = 0.80$ for cluster~4
and $S_\hash = 0.37$ and $S_\dir = 0.96$ for cluster~5.  For cluster~4, the
matching versions were also in quite a tight range from September 2013 to
September 2014 (our versions 9 to 11), whereas most other clusters ranged
more evenly across all 18 versions.

\item{\textbf{6 and 7.}} These clusters consisted largely of forks from
Litecoin: 100\% of cluster~6 had the file \verb#scrypt.c#, which is unique to
Litecoin, and indeed 100\% identified as Litecoin derivatives using the
copyright method from Section~\ref{sec:copyright}.  64\% of
cluster~7 had files with \verb#scrypt# in the name, although only 21\%
identified as copyright derivatives of anything other than Bitcoin.

\item{\textbf{8.}} The nodes in this cluster were on average newer than the
others (with the first repository created in June 2015), and indeed their
directory structure is more consistent with newer versions of the Bitcoin
codebase.

\end{itemize}

\section{Ethereum as a Platform}\label{sec:ethereum}

As discussed in Sections~\ref{sec:back} and~\ref{sec:contracts}, it is
increasingly popular to deploy cryptocurrencies as tokens on the
Ethereum blockchain;
indeed, over half of the cryptocurrencies listed on CoinMarketCap fell into
this category.  This section thus explores this type of cryptocurrency
deployment, focusing again on the extent to which ERC20 tokens are similar to
or different from each other.  As an ERC20 token consists of just a single file,
our methods from the previous sections do not apply here so we develop new
methods for identifying similarities.
The basic functionality of an ERC20 token\dash allowing the transfer of tokens
from one holder to another\dash defines a contract type called \verb#Basic# (or
\verb#BasicToken#) or\dash with one slight functional difference\dash
\verb#ERC20#.
There are, however, many additional types that ERC20 tokens can have.
For example, if they want
to allow for the creation of new tokens they can be \verb#Mintable#
and if they want to allow for the destruction of existing tokens
they can be \verb#Destructible# or \verb#Burnable#.  These types are not
standardized, and in fact new types can be defined and used within the
Solidity code for a contract.  %

To identify the types of a given token, we identified all lines in its
contract of the form
\verb#contract X is Y {#, where \verb#X# is the name of the contract and
\verb#Y# is its type.  Some intermediate types themselves appear as names
(e.g., \verb#contract Mintable is Ownable#), which we exclude from our final
results but carry over transitively to the higher-level contract names; e.g.,
if \verb#X# is \verb#Mintable# and
\verb#Mintable# is \verb#Ownable# then \verb#X# is both \verb#Mintable# and
\verb#Ownable#.  This resulted in a map from the higher-level token
names to a list of all of their types.

Beyond these types, there are several ERC20 templates whose reuse
we sought to identify.  Based on a manual inspection of a
random subset of contracts, we chose five points of distinction: first,
which version the contract used of (1)
Solidity and (2) the SafeMath library, which provides safe
arithmetic operations.  Second, we considered whether or
not it used (3) the definition of \verb#StandardToken# created by the
FirstBlood token; (4) the \verb#UpgradeableToken# type, created by
the Golem token; or (5) a template by
OpenZeppelin,\footnote{\url{https://openzeppelin.org/}} who also created
SafeMath.

For the version of Solidity, we looked for lines starting with
\verb#pragma solidity# and extracted the version from what followed
(typically of the form \verb#0.4.X#).  To determine the version of SafeMath,
we first used CLOC to strip the comments from the \verb#.sol# file.  We
then identified the lines of code that defined the SafeMath library
(starting with either \verb#contract SafeMath {# or
\verb#library SafeMath {# and ending with \verb#}#), and hashed this substring
to form a succinct representation.  Finally, to identify the use of templates,
we simply looked at whether or not the file explicitly mentioned the relevant
keywords.

\begin{figure}[t]
\centering
\begin{subfigure}[b]{0.295\textwidth}
\centering
\includegraphics[width=\textwidth]{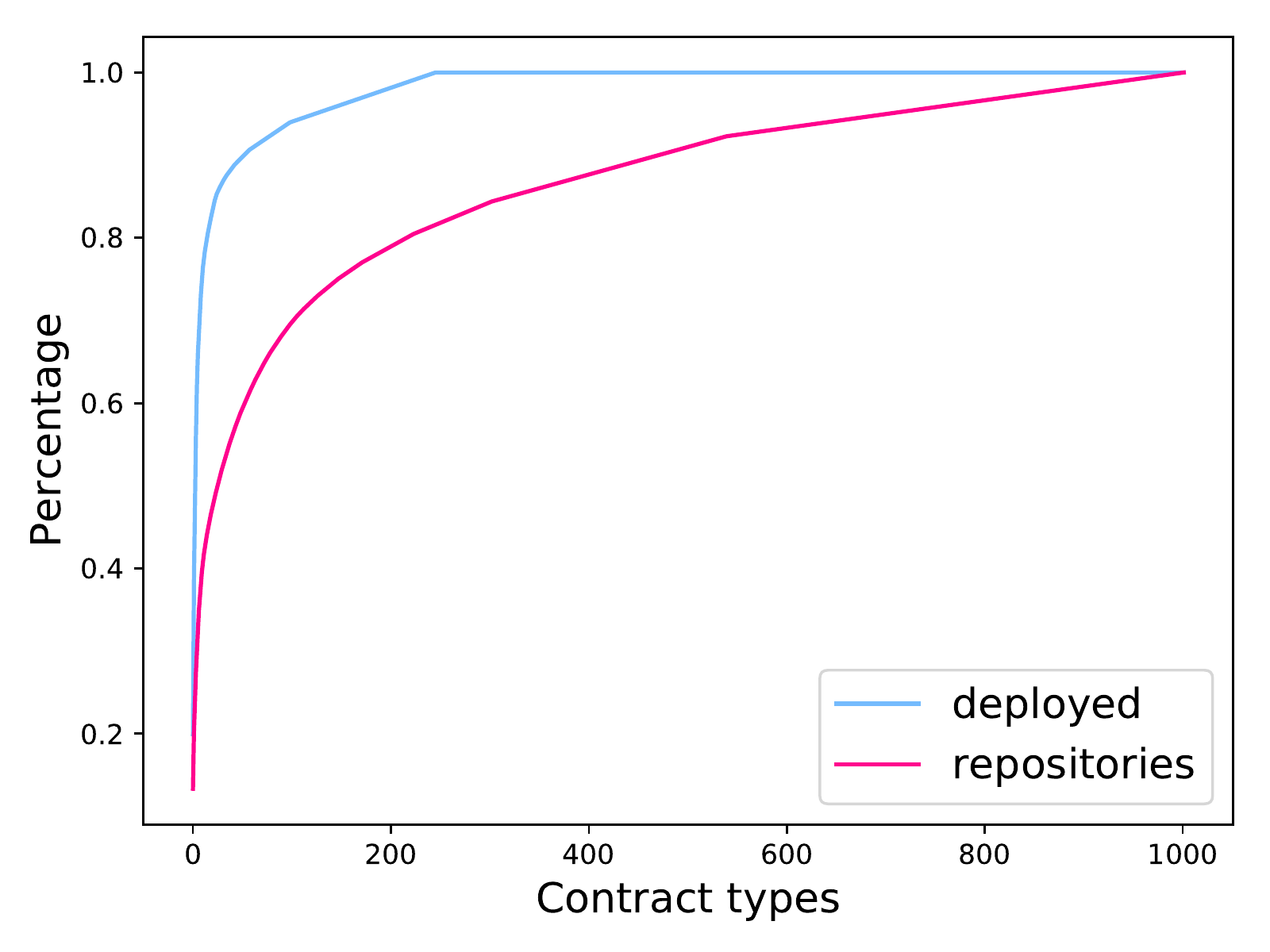}
\caption{Types}
\label{fig:types}
\end{subfigure}
~
\begin{subfigure}[b]{0.32\textwidth}
\centering
\includegraphics[width=\textwidth]{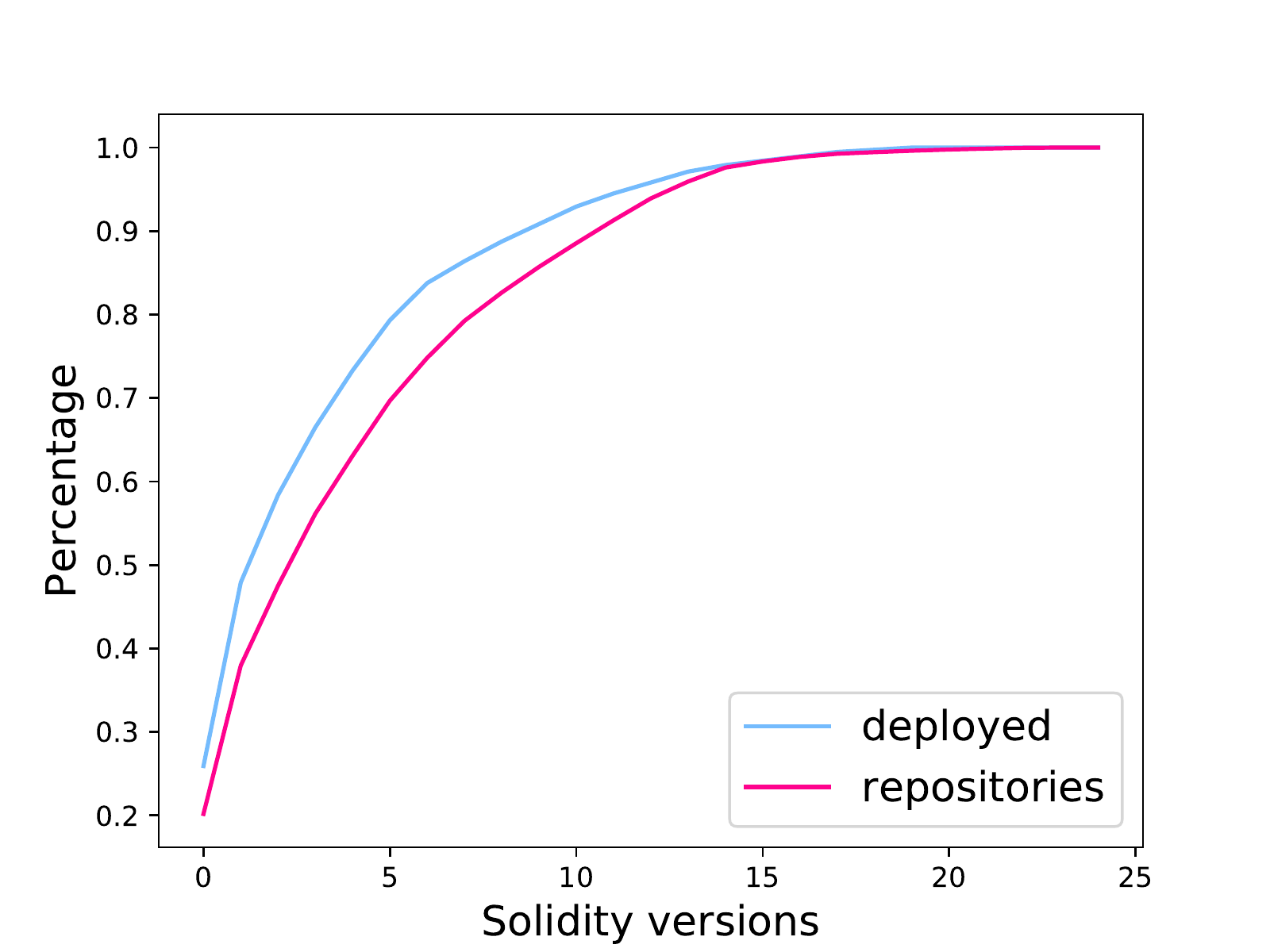}
\caption{Solidity version}
\label{fig:version}
\end{subfigure}
~
\begin{subfigure}[b]{0.32\textwidth}
\centering
\includegraphics[width=\textwidth]{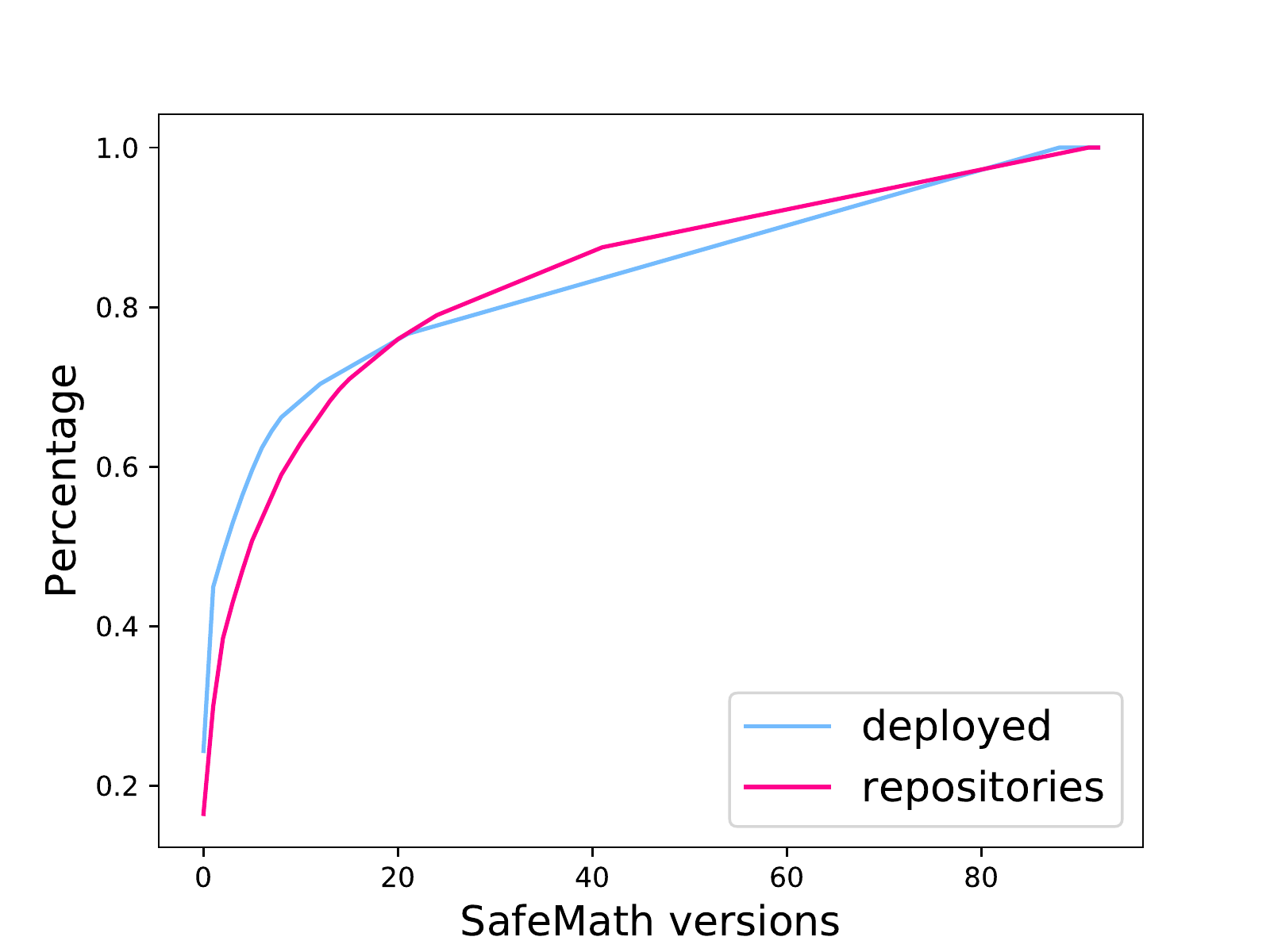}
\caption{SafeMath version}
\label{fig:safemath}
\end{subfigure}
\caption{When ranked from most to least popular, the cumulative percentage of
contracts matching three different features, for both the set of deployed
contracts and the ones found in repositories.}
\label{fig:stuff-cdf}
\hrulefill
\end{figure}

We extracted this information from all Solidity files, whether deployed on
the Ethereum blockchain (and thus scraped from Etherscan, as
described in Section~\ref{sec:contracts}) or contained in a
repository.\footnote{Interestingly, these sets were non-intersecting; i.e.,
    there was no contract in a repository that was identical to a deployed one.}
For the types, Solidity and SafeMath versions, we ordered them from
most to least popular and plotted this as a CDF, as seen in
Figure~\ref{fig:stuff-cdf}; i.e., we plotted the percentage $y$ of all
contracts that had one of the top $x$ attributes.  For the templates, the
results are in Table~\ref{tab:contract-stuff}.

\begin{table}[t]
\setlength{\tabcolsep}{8pt}
\centering
\begin{tabular}{lS[table-format=3]S[table-format=2.1]S[table-format=3]S[table-format=2.1]}
\toprule
& \multicolumn{2}{c}{Deployed} & \multicolumn{2}{c}{Repositories} \\
\cmidrule(lr){2-3} \cmidrule(lr){4-5}
& {\#} & {\%} & {\#} & {\%} \\
\midrule
FirstBlood & 134 & 30.8 & 190 & 2.5 \\
Lunyr & 22 & 5.1 & 19 & 0.2 \\
OpenZeppelin & 72 & 16.6 & 245 & 3.2 \\
SafeMath & 287 & 65.9 & 400 & 5.2 \\
\bottomrule
\end{tabular}
\caption{The number of contracts of each type (deployed or in repositories)
that are derived from one of the first three templates, or using the SafeMath
library, as well as the percentage of all contracts this represents.}
\label{tab:contract-stuff}
\hrulefill
\end{table}

The relatively long tails in all of the figures indicate a relatively high level
of diversity among these features in both deployed contracts and those still
under development.  For example, the Solidity version most popular among
deployed contracts (version~18) was still used in only 23\% of them.
Whereas Figures~\ref{fig:version}
and~\ref{fig:safemath} show similar curves for both sets of contracts,
Figure~\ref{fig:types} shows a much longer tail for contracts contained in
repositories, with 246 distinct types in deployed contracts and 1002 in ones
in repositories.  This indicates\dash as should perhaps be expected\dash that
(1) there are just many more possibilities for contract types than for versions,
and (2) there is greater experimentation with types in
contracts still under development.  Even among deployed contracts, 129 out of
429 had a type that did not appear in any other
deployed contracts, and 148 of the 246 distinct types appeared in only a single
contract.
\sarah{Worth giving the same stats for repo contracts too?}

Table~\ref{tab:contract-stuff} also demonstrates the relatively high
diversity across contracts, with no one template being used in a dominant
way.  Even though 65.9\% of deployed contracts use SafeMath,
Figure~\ref{fig:safemath} demonstrates that there is
quite a lot of variety within this category; indeed, there were 90 different
versions of SafeMath used in deployed contracts (and 93 different versions
in the repositories).  Again, we see significantly more experimentation in
contracts still under development.

Finally, we view the points of similarity that did exist as operating
primarily in support of the safety of deployed contracts.  For example,
among the 20 most popular types across both
deployed and repository contracts, five of them defined the basic
ERC20 functionality,
and six of them were related to safety in terms of either
including a standard library %
or in defining an owner
who could take action
if something went wrong.  The same is
true of the usage of FirstBlood's \verb#StandardToken#, which was the first
safe implementation of this type, or of the \verb#SafeMath# library.  We thus
view these similarities as a sign of good development practices, rather than
the copying of ideas.

\section{Conclusions}\label{sec:conclusions}

This paper considered diversity in the cryptocurrency landscape, according to
the source code available for each one, in order to identify the extent to
which new cryptocurrencies provide meaningful innovation.  This was
done by examining the source code for over a thousand cryptocurrencies,
and\dash in the case of ERC20 tokens\dash the deployed code of hundreds more.
While more sophisticated static analysis of the source code would likely
yield further insights, even our relatively coarse methods clearly indicated
the dominance of Bitcoin and Ethereum, as well
as the extent to which creating a standalone platform is a significantly
greater undertaking (leading to the reuse of much of the Bitcoin codebase)
than defining just the transaction semantics of an Ethereum-based token.

\section*{Acknowledgements}

The authors were supported in part by EPSRC Grant EP/N028104/1 and in part by
the EU H2020 TITANIUM project under grant agreement number 740558.

{\footnotesize
\bibliographystyle{abbrv}
\bibliography{abbrev2,crypto,misc}
}

\appendix

\section{Extra Tables and Figures}\label{sec:extras}

\exclusiontable

\bitcoindir

\end{document}